\documentclass[prl, twocolumn, superscriptaddress]{revtex4-1}

\usepackage{graphicx,subfigure}
\usepackage{dcolumn}
\usepackage{epsfig}
\usepackage{latexsym}
\usepackage{amsfonts}
\usepackage{amssymb}

\begin{document}

\title{Elementary edge and screw dislocations visualized at the 
	lattice periodicity level 
	in smectic phase of colloidal rods \\
	}

\author{Andrii Repula$^{1}$ and Eric Grelet}
\altaffiliation{grelet@crpp-bordeaux.cnrs.fr}
\affiliation{
Centre de Recherche Paul-Pascal, CNRS \& Universit\'e de Bordeaux, 115 Avenue Schweitzer, F-33600 Pessac, France \\
}

\date{\today}
\begin{abstract}
	 
\textbf{Compelling justification:} Topological defects such as dislocations play a major role in science, from condensed matter and geophysics to cosmology. Line defects 
in periodically ordered structures, called dislocations, mediate phase transitions and determine many distinctive features of materials, from crystal growth to mechanical properties. However, despite many theoretical predictions, the 
detailed structure of dislocations remains largely unexplored. By using a model system of tip-labeled colloidal rod-shaped particles 
enabling improved resolution and contrast by optical microscopy,
\textit{in situ} visualization of dislocations has been performed at the lattice periodicity 
level in a self-organized layered phase. 
The local morphology 
of dislocations has been quantitatively determined, 
evidencing experimentally a ``melted'' defect core. 
\\

\textbf{Abstract:} We report on the identification and quantitative characterization  
of elementary edge and screw dislocations in a colloidal smectic phase of tip-labeled rods. 
Thanks to the micrometer layer spacing, direct visualization of dislocations has been performed at the \textit{smectic periodicity scale} by optical fluorescence microscopy. As a result, the displacement field around an edge dislocation has been experimentally established and compared with the profile
predicted by elastic theory. 
Elementary screw dislocations have been also evidenced, for which the core size as well as the \textit{in situ} handedness have been determined. 
Self-diffusion experiments performed at the individual particle level reveal for the first time nematic-like or ``melted'' ordering of the defect core.  

\end{abstract}


\maketitle

\begin{figure}
	\includegraphics[width=1\columnwidth]{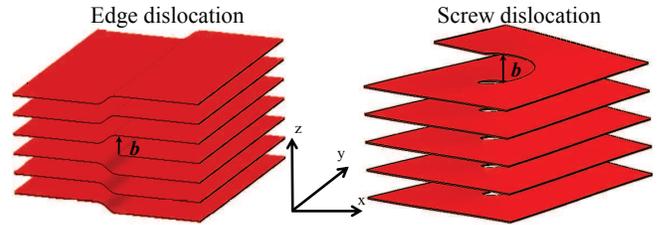}
	\caption{Schematic representations of an edge and a screw dislocations in a lamellar system. The Burgers vector \textit{b} is perpendicular to the dislocation line along $y$ for an edge dislocation and it is parallel to the dislocation line along $z$ in a screw dislocation. 
	}
	\label{Schemes}
\end{figure}

Topological defects are ubiquitous in all ordered systems, and are generated either extrinsically as 
in plastic deformation, or intrinsically in frustrated materials such as liquid crystalline blue phases \cite{Chaikin-Lubensky,Kleman2004,Kleman-Lavrentovich,Oswald-Pieranski}. Recently, they have been widely used to manipulate micro- and nanoparticles in different mesophases \cite{Iwashita2003,Tkalec2011,Coursault2012,Abbott2016}. Among topological singularities, dislocations are linear defects with broken translational symmetry, and they determine many properties of regular solids \cite{Kleman1983,Hanschke2017}, including important mechanical ones in metals \cite{Friedel1964}. In systems with reduced dimensionality, such as layered phases usually found in soft condensed matter, dislocations exhibit substantial differences compared with their 3D counterparts \cite{Kleman1983,Holyst1995,Oswald-Pieranski,Pieranski2016}. Despite their major role in liquid crystals \cite{Meyer1978,Kamien1997,Trivedi2012,Kamien2017} and in recently evidenced colloidal self-assembly where dislocations have been shown to mediate chirality transfer from constituent particles to helical superstructures \cite{NatCom}, quantitative experimental characterizations of the dislocation displacement field as well as the core structure of elementary dislocations are scarce to date \cite{Oswald-Pieranski}. Such investigations rely on the observation of dislocations at length scales corresponding to the lamellar peridiocity, which is of a few nanometers for most thermotropic, amphiphilic and block copolymer based liquid crystals \cite{Holyst1995,Chan1981,Maloum1992}. This therefore usually requires electron microscopy techniques to gain access to nanometer resolved images, combined with a control of the sample alignment \cite{Zasadzinski1990,Zhang2015}. If such difficulties have been partially bypassed by studying by optical microscopy some cholesteric mesophase whose fingerprint texture is reminiscent of a lamellar ordering \cite{Ishikawa,Smalyukh2002,Engstrom2011}, the detailed structure of elementary dislocations in real smectics is not yet fully achieved \cite{Zhang2013,Zhang2015,Kamien2017}. Specifically, the local order within the dislocation core is expected to be ``melted'' into a higher-symmetry phase; however, the nature of this order, either liquid-like or nematic-like, did not receive yet a full experimental confirmation. In this work, we have successfully produced tip functionalized rod-like particles, namely fd viruses fluorescently labeled at one end, whose monodisperse micrometer length results in the formation of colloidal smectic phase at high enough volume fractions. Taking advantage of its micrometer periodicity which enables its study by optical microscopy, the tip-labeled virus based colloidal smectic phase with planar alignment is found to exhibit elementary dislocations, namely edge and screw ones, as schematically represented in Fig. \ref{Schemes}.  
                 
\begin{figure}
	\includegraphics[width=0.95\columnwidth]{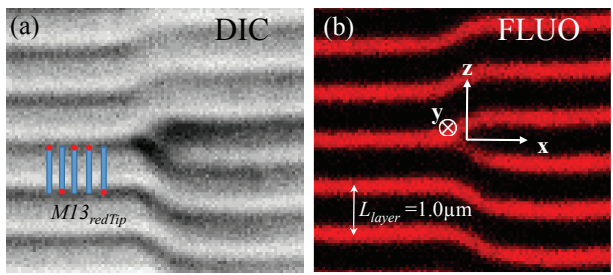}
	\caption{Imaging of smectic layers and an edge dislocation in suspensions of tip-functionalized colloidal rods ($M13_{redTip}$, schematically represented by blue dashes with red dots) observed respectively by (a) differential interference contrast (DIC) and (b) fluorescence optical microscopy. 
		The length scale is provided by the micrometer layer spacing, $L_{layer}$. The origin for profile measurement is located at the dislocation core. In the observation plane, the $x$ and $z$ axes are parallel and perpendicular to the layers, while the focal depth is along the $y$ axis.  }
	\label{Edge-DIC}
\end{figure}

\begin{figure}
	\begin{center}
		\includegraphics[width=0.65\columnwidth]{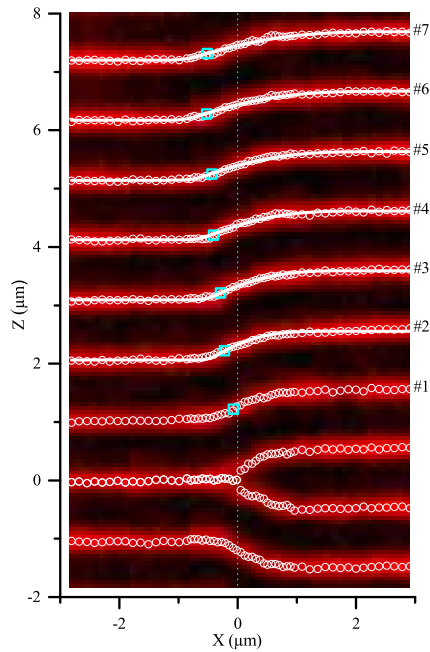}	
	\end{center}
	\caption{Edge dislocation observed by fluorescence microscopy and the corresponding displacement field (white open circles) of the adjacent smectic layers around the defect core. The white solid lines represent the global fitting of the dislocation profile according to the nonlinear elastic theory (Eq. \ref{Non-Linear}) from which the elastic length $\lambda$ (Eq. \ref{lambda}) can be deduced. The inflection points are marked by blue open squares, and are all shifted toward the same direction ($x<0$) as predicted by the theory (Eq. \ref{Non-Linear}). 
	}
	\label{EdgeDisloc}
\end{figure} 

Experimentally, we used 
M13KE filamentous viruses, as system of rod-like particles. 
Each of the five P3 proteins localized at the virus proximal end 
displays pairs of cysteine residues (otherwise absent from the other coat proteins of the virus), forming disulfide bridges. After reduction by Tris(2-carboxyethyl)phosphine hydrochloride (TCEP, Thermo Scientific), the resulting thiol groups are available for conjugation with maleimide activated compounds. We chose to specifically label these thiol groups present at the virus tip with red dyes (DyLight™ 550 Maleimide, ThermoFisher) according to a  protocol reported elsewhere \cite{ACSnano}. This results in $1.0~\mu m$ long tip-labeled viral particles, called $M13_{redTip}$, whose phase behavior in aqueous suspension presents liquid crystalline ordering, as the lamellar organization including smectic-A and smectic-B phases in the dense regime \cite{Dogic2006,Grelet2014}. Part of these tip-functionalized viruses were body-labeled with green fluorescent dyes (Alexa488-NHS ester activated, Molecular
Probes) \cite{ACSnano} and added in tracer amounts to a few $M13_{redTip}$ samples for single particle tracking experiments.   
A set of samples with concentrations in the smectic-A range were prepared (in 17mM BisTris-HCl-NaCl buffer, pH 7, with 3~mM NaN$_3$ as bactericide, for a total ionic strength of 20~mM) between cover slip and glass slides, and observations were performed using an optical microscope (IX-71 Olympus), equipped with a high-numerical aperture (NA) oil objective (100X PlanApo NA 1.40), a piezo device for objective vertical positioning (P-721 PIFOC Piezo Flexure Objective
Scanner, PI), a fluorescence excitation light source (X-cite series 120 Q) and an ultra-fast electron-multiplying camera (NEO sCMOS Andor) with a pixel size of 6.5~$\mu m$. An equilibrium time of typically a few days was applied in order to get homogeneous and uniform smectic-A samples. This could result sometimes in a partial drying of the samples, leading possibly to some smectic-B ordering.  

The resulting smectic phase formed by the tip-labeled rod-like viruses, is observed both by differential interference contrast and fluorescence microscopy, as shown in Figure \ref{Edge-DIC}.
Both contrast modes reveal the lamellar periodicity of the smectic organization, with improved resolution and contrast in case of fluorescence microscopy. As the tips of our colloidal rods have been labeled with red dyes, the smectic layer positions as well as the layer spacing, $L_{layer}$, can be experimentally determined with a good level of precision. This allows for the quantitative analysis of the displacement field around an elementary edge dislocation, as reported in Fig. \ref{EdgeDisloc}. The general expression of the nonlinear elastic free energy density of a smectic phase is, by accounting for the higher order derivatives in the displacement field, $u(x,z)$ \cite{Kleman-Lavrentovich}:
\begin{equation}
f=\frac{B}{2}
\left(
\frac{\partial u}{\partial z}
- \frac{1}{2}
\left(\frac{\partial u}{\partial x}\right)^2
\right)^2
+ \frac{K_1}{2}
\left(\frac{\partial^2 u}{\partial x^2}\right)^2 
\label{freeEnergy}
\end{equation}

with $B$ the compressibility modulus and $K_1$ the bending elastic constant, analogous to the nematic splay elastic constant. The elastic length $\lambda$ can then be defined as: 
\begin{equation}
\lambda=\sqrt{\frac{K_1}{B}}
\label{lambda}
\end{equation}  

providing the intrinsic length scale over which an imposed distortion relaxes in a lamellar system.
Using Eq. \ref{freeEnergy}, Brener and Marchenko found the analytical edge dislocation profile \cite{Brener99}:
\begin{equation}
u(x,z)=2\lambda \ln \left[
1+\frac{\exp(b/4\lambda)-1}{2}
\left(1+erf\left(\frac{x}{2\sqrt{\lambda z}}\right)
\right)\right]
\label{Non-Linear}
\end{equation}   
where $erf(...)$ is the error function. In the limit of small Burgers vector $b\ll\lambda$, Eq. \ref{Non-Linear} reduces to the classical result of the linear elastic theory \cite{DeGennes1972,Kleman1983}: 
\begin{equation}
u(x,z)=\frac{b}{4}
\left[1+erf\left(\frac{x}{2\sqrt{\lambda z}}\right)\right]
\label{Linear}
\end{equation}  

Figure \ref{EdgeDisloc} shows the measured displacement field of the smectic layers around a dislocation core. Note that it is difficult experimentally to have fully isolated edge dislocations: most are disturbed by elastic distortion stemming either from the presence of other defects and grain boundaries, or from heterogeneous anchoring at the cell walls, 
which can result in a lack of mirror symmetry of the defect.  
As expected for an edge dislocation of Burgers vector $b/L_{layer}=1$, the total displacement of a given layer is $b/2$ regardless the layer distance from the defect core. This has been experimentally checked within an accuracy greater than 95\% for the 6 considered layers (Fig. \ref{EdgeDisloc}), proving \textit{a posteriori} that the chosen edge dislocation is nearly isolated. This is also confirmed from the quantitative analysis of the dislocation profile, performed according to Eq. \ref{Non-Linear}. 
The smectic layers of interest have been globally fitted with a single value of the elastic length $\lambda$ to account for the overall displacement field.
With a resulting value of $\lambda = 0.02\pm0.01\mu m$, a very good agreement with the nonlinear elastic theory is found far enough from the dislocation core (from layers \#3 to \#7, Fig. \ref{EdgeDisloc}). The first layers, which are too close to the defect core, have been either disregarded from the quantitative analysis (layer \#1), or are approximatively fitted by the model (layer \#2). As $\lambda \ll b$, only the nonlinear elastic theory is expected to apply, with some intrinsic features clearly evidenced in our experimental data. This includes a shift of the inflection points towards the smectic region with the missing layer ($x<0$, Fig. \ref{EdgeDisloc}) and an asymmetry of the dislocation profile with a sharp rise at $x<0$ and a smooth saturation at $x>0$ \cite{Brener99,Ishikawa}.
It is worth mentioning that if the nonlinear elastic theory (Eq. \ref{Non-Linear}) is able to account for the experimental displacement field, the quantitative comparison  remains quite sensitive to any extrinsic disturbance of the defect profile. 


From the determination of the elastic length $\lambda$, the compressibility modulus $B$ can be calculated knowing the bending modulus $K_1$ (Eq. \ref{lambda}). In fluid-like membranes composed of one-rod length thick monolayer of aligned viruses, the bending modulus has been found to be $\kappa _c \simeq 150~k_B T$ \cite{Barry2010}. To extend this value found for a single membrane to a bulk smectic phase composed of similar colloidal viral rods, we have to renormalize $\kappa _c$ by the smectic layer spacing $L_{layer}$ to give an estimation of the 3D bending modulus $K_1=\kappa _c/L_{layer}\approx150~k_B T / \mu m$. It is worth pointing out that this value is in good agreement with the value obtained from the one-Frank constant approximation $K_1\approx K_2 = 0.5~pN = 100~k_B T / \mu m$, where $K_2$ is the twist elastic constant measured by unwinding the cholesteric phase of virus suspensions under magnetic field \cite{Dogic2000}. Knowing $K_1$ and $\lambda$, we can therefore estimate, according to Eq. \ref{lambda}, the smectic compressibility modulus $B\simeq 4\times10^5 ~k_B T / \mu m ^3$  for a smectic phase of colloidal rods. This value can be compared with the typical one found in thermotropic smectics \cite{Shibahara2000}, $B_{thermotropic}\sim 10^8 ~dyn / cm^2\simeq 2\times10^9 ~k_B T / \mu m ^3 = 2 ~k_B T / nm ^3$. This means that if the absolute value of the compressibility modulus is higher by a few orders of magnitude in thermotropic liquid crystals, the same value rescaled by the molecular length (nanometers in size in thermotropics compared with the micrometer length for viral colloidal rods) indicates a softer rescaled layer compressibility in molecular systems.  


\begin{figure}
	\begin{center}
		\includegraphics[width=1.0\columnwidth]{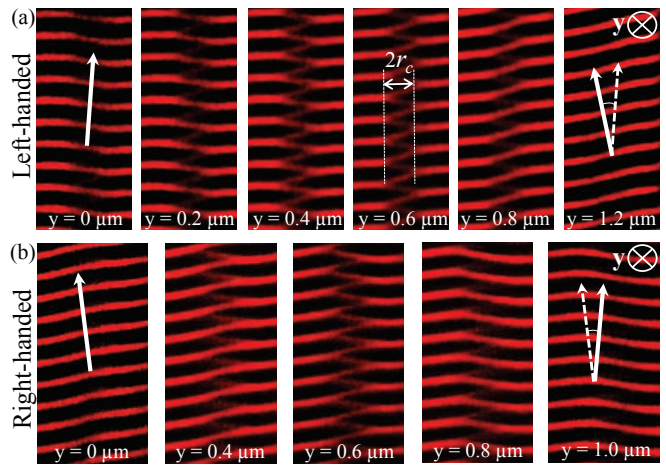}	
	\end{center}
	\caption{Screw dislocations with opposite handedness observed by fluorescence microscopy in smectic phase of tip-labeled viruses. The handedness has been determined by capturing images at different focal depths $y$, resulting in (a) left- and (b) right-handed screw dislocations, as indicated by the rotation of the normal to the smectic layers shown by the white arrows. The length scale is provided by the 1~$\mu m$ long smectic layer spacing. }
	\label{ScrewDisloc}
\end{figure} 

\begin{figure*}
	\begin{center}
		\includegraphics[width=1.2\columnwidth]{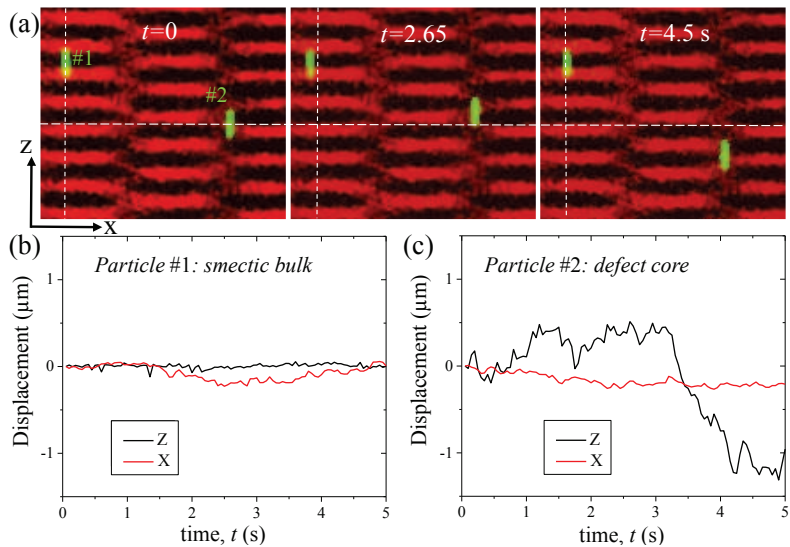}
	\end{center}
	\caption{(a) Overlays at different times of fluorescence microscopy images showing red smectic layers from $M13_{redTip}$ particles, in which tracer amounts of $M13_{redTip}$ particles body-labeled with green fluorescent dyes have been added, enabling single particle tracking experiments. The vertical dotted and horizontal dashed lines are visual guides to emphasize the displacements of green viral particles \#1 and \#2, respectively. The length scale  is provided by the 1~$\mu m$ long smectic layer spacing. (b) Example of a trajectory within the smectic-A bulk where the particle mainly displays lateral motion within the layers. 
	(c) Trajectory of a viral rod-like particle within the defect core evidencing its rapid parallel (i.e. along the $z$ axis) diffusion, characteristic of nematic-like behavior.  
	}
	\label{Dynamics}
\end{figure*} 

In our smectic samples of tip-labeled-viruses, only elementary dislocations have been observed, with a Burgers vector magnitude $b~=1\times L_{layer}=1~\mu m$ (Fig. \ref{Schemes}) whatever the dislocation type (either edge type or screw type). Examples of screw dislocations are shown in Fig. \ref{ScrewDisloc}, 
where, remarkably, their core extension can be clearly distinguished. This allows for the \textit{in situ} handedness determination by varying the optical microscope focus through the sample thickness, $y$, resulting in the presence of both left-handed and right-handed screw dislocations (Fig. \ref{ScrewDisloc}), as expected in an achiral smectic-A mesophase. Here, we benefit from the micrometer length scale of our experimental system, 
enabling structural investigations that are otherwise very tricky to achieve using usual molecular nanometric liquid crystals \cite{Zhang2015}.
Specifically, the screw dislocation core size $2r_c$ can be measured, as indicated in Fig. \ref{ScrewDisloc}. The distribution of core radii has been determined and is shown in the Supplemental Material \cite{SM}: a mean value of $2r_c = 0.98~\mu m \simeq b$ is found. This value of the screw dislocation core size is in good agreement with the one determined by Zhang \textit{et al}. \cite{Zhang2013}, but is larger than the theoretical prediction $2r_c \approx b/\pi$ performed by Pleiner \cite{Pleiner1986,Pleiner1988}. Theoretical models should therefore be developed for a more detailed description of the microstructure of the screw dislocation core \cite{Matsumoto12} to account for the experimental observations.  
The introduction of tracer amounts of additionally body-labeled viruses with green fluorescent dyes in the smectic samples provides further information about the dislocation core. 
Figure \ref{Dynamics} reveals the dynamics at the single particle level, and a main difference distinguishes the self-diffusion within the dislocation core, where a body-labeled rod has been trapped, from the smectic bulk: 
contrary to viral particles in the smectic bulk which diffuse slightly within the layers with rare hopping type events \cite{Alvarez2017} (See the Supplemental Material \cite{SM}), 
the diffusion of the particles within the defect core is strongly enhanced, with a large displacement, for a given observation time, 
along the director (or, equivalently, along the normal to the smectic layers indicated by the $z$
axis). This dynamics with a motion mainly along the long rod axis and without significant rotational diffusion, is characteristic of nematic-like behavior \cite{Alvarez2017}, and represents to the best of our knowledge, the first demonstration of the nematic ordering associated with the higher-symmetry phase forming the dislocation core in smectic defects. This nematic order is in qualitative agreement with phenomenological Landau-de Gennes type modeling of screw dislocation core structures \cite{Pleiner1986,Kralj1993}, and our experimental results therefore rule out dislocation models based on either isotropic liquid core or strongly distorted smectic layers at the vicinity of the defect core \cite{Pleiner1986,Pleiner1988,Kralj1993,Kralj1995}.  
Despite the larger number of screw dislocations observed in our samples compared to edge ones, which is consistent with their respective free energy \cite{Kleman1983,Kleman-Lavrentovich}, 
it turns out that only a very few labeled particles have been seen in the defect core (See the Supplemental Material \cite{SM}), making further quantitative analysis of the dynamics of this system difficult.

To conclude, we have identified and quantitatively characterized the microscopic structure of edge and screw dislocations in a smectic phase of functionalized colloidal rod-like particles. Thanks to a regioselective labeling of the filamentous virus tips by fluorescent dyes, high resolution and contrast imaging of the topological defects has been achieved by fluorescence microscopy, revealing the displacement field around elementary edge dislocation, as well as some unprecedented detailed information of the screw dislocation core, such as its size, helical handedness and local structure with a nematic-like ordering. A perspective of our work is the use of topological defects formed by \textit{colloidal liquid crystals} to organize nano- and microparticle assemblies. 

\begin{acknowledgments}
This project has received funding from the European
Union Horizon 2020 research and innovation programme
under the Marie Sk\l{}odowska-Curie Grant Agreement No.
641839. We thank A. de la Cotte for preliminary experiments, C. Wu for help in sample preparation, P. Merzeau for drawing the schemes, and F. Nallet for discussions. \\
Author contributions: A.R. prepared the samples and performed experiments, E.G. analyzed the data and wrote the paper.
\end{acknowledgments}

\bibliography{Disloc}	

\begin{thebibliography}{41}
\expandafter\ifx\csname natexlab\endcsname\relax\def\natexlab#1{#1}\fi
\expandafter\ifx\csname bibnamefont\endcsname\relax
  \def\bibnamefont#1{#1}\fi
\expandafter\ifx\csname bibfnamefont\endcsname\relax
  \def\bibfnamefont#1{#1}\fi
\expandafter\ifx\csname citenamefont\endcsname\relax
  \def\citenamefont#1{#1}\fi
\expandafter\ifx\csname url\endcsname\relax
  \def\url#1{\texttt{#1}}\fi
\expandafter\ifx\csname urlprefix\endcsname\relax\def\urlprefix{URL }\fi
\providecommand{\bibinfo}[2]{#2}
\providecommand{\eprint}[2][]{\url{#2}}

\bibitem[{\citenamefont{Chaikin and Lubensky}(1995)}]{Chaikin-Lubensky}
\bibinfo{author}{\bibfnamefont{P.~M.} \bibnamefont{Chaikin}} \bibnamefont{and}
  \bibinfo{author}{\bibfnamefont{T.~C.} \bibnamefont{Lubensky}},
  \emph{\bibinfo{title}{Principles of Condensed Matter Physics}}
  (\bibinfo{publisher}{Cambridge Univ Press}, \bibinfo{address}{Cambridge},
  \bibinfo{year}{1995}).

\bibitem[{\citenamefont{Kleman et~al.}(2004)\citenamefont{Kleman, Lavrentovich,
  and Nastishin}}]{Kleman2004}
\bibinfo{author}{\bibfnamefont{M.}~\bibnamefont{Kleman}},
  \bibinfo{author}{\bibfnamefont{O.~D.} \bibnamefont{Lavrentovich}},
  \bibnamefont{and} \bibinfo{author}{\bibfnamefont{Y.~A.}
  \bibnamefont{Nastishin}}, in \emph{\bibinfo{booktitle}{Dislocations in
  Solids}}, edited by \bibinfo{editor}{\bibfnamefont{F.~R.~N.}
  \bibnamefont{Nabarro}} \bibnamefont{and}
  \bibinfo{editor}{\bibfnamefont{J.~P.} \bibnamefont{Hirth}}
  (\bibinfo{publisher}{Elsevier}, \bibinfo{address}{Amsterdam Vol. 12, p. 147},
  \bibinfo{year}{2004}).

\bibitem[{\citenamefont{Kleman and Lavrentovich}(2002)}]{Kleman-Lavrentovich}
\bibinfo{author}{\bibfnamefont{M.}~\bibnamefont{Kleman}} \bibnamefont{and}
  \bibinfo{author}{\bibfnamefont{O.~D.} \bibnamefont{Lavrentovich}},
  \emph{\bibinfo{title}{Soft Matter Physics: An Introduction}}
  (\bibinfo{publisher}{Springer-Verlag}, \bibinfo{address}{New York},
  \bibinfo{year}{2002}).

\bibitem[{\citenamefont{Oswald and Pieranski}(2005)}]{Oswald-Pieranski}
\bibinfo{author}{\bibfnamefont{P.}~\bibnamefont{Oswald}} \bibnamefont{and}
  \bibinfo{author}{\bibfnamefont{P.}~\bibnamefont{Pieranski}},
  \emph{\bibinfo{title}{Smectic and Columnar Liquid Crystals: Concepts and
  Physical Properties Illustrated by Experiments}} (\bibinfo{publisher}{Taylor
  and Francis/CRC Press}, \bibinfo{address}{London}, \bibinfo{year}{2005}).

\bibitem[{\citenamefont{Iwashita and Tanaka}(2003)}]{Iwashita2003}
\bibinfo{author}{\bibfnamefont{Y.}~\bibnamefont{Iwashita}} \bibnamefont{and}
  \bibinfo{author}{\bibfnamefont{H.}~\bibnamefont{Tanaka}},
  \bibinfo{journal}{Phys. Rev. Lett.} \textbf{\bibinfo{volume}{90}},
  \bibinfo{pages}{045501} (\bibinfo{year}{2003}).

\bibitem[{\citenamefont{Tkalec et~al.}(2011)\citenamefont{Tkalec, Ravnik,
  \u{C}opar, \u{Z}umer, and Mu\u{s}evi\u{c}}}]{Tkalec2011}
\bibinfo{author}{\bibfnamefont{U.}~\bibnamefont{Tkalec}},
  \bibinfo{author}{\bibfnamefont{M.}~\bibnamefont{Ravnik}},
  \bibinfo{author}{\bibfnamefont{S.}~\bibnamefont{\u{C}opar}},
  \bibinfo{author}{\bibfnamefont{S.}~\bibnamefont{\u{Z}umer}},
  \bibnamefont{and}
  \bibinfo{author}{\bibfnamefont{I.}~\bibnamefont{Mu\u{s}evi\u{c}}},
  \bibinfo{journal}{Science} \textbf{\bibinfo{volume}{333}},
  \bibinfo{pages}{62} (\bibinfo{year}{2011}).

\bibitem[{\citenamefont{Coursault et~al.}(2012)\citenamefont{Coursault, Grand,
  Zappone, Ayeb, L{\'e}vi, F{\'e}lidj, and Lacaze}}]{Coursault2012}
\bibinfo{author}{\bibfnamefont{D.}~\bibnamefont{Coursault}},
  \bibinfo{author}{\bibfnamefont{J.}~\bibnamefont{Grand}},
  \bibinfo{author}{\bibfnamefont{B.}~\bibnamefont{Zappone}},
  \bibinfo{author}{\bibfnamefont{H.}~\bibnamefont{Ayeb}},
  \bibinfo{author}{\bibfnamefont{G.}~\bibnamefont{L{\'e}vi}},
  \bibinfo{author}{\bibfnamefont{N.}~\bibnamefont{F{\'e}lidj}},
  \bibnamefont{and} \bibinfo{author}{\bibfnamefont{E.}~\bibnamefont{Lacaze}},
  \bibinfo{journal}{Adv. Mater.} \textbf{\bibinfo{volume}{24}},
  \bibinfo{pages}{1461} (\bibinfo{year}{2012}).

\bibitem[{\citenamefont{Wang et~al.}(2016)\citenamefont{Wang, Miller,
  Bukusoglu, de~Pablo, and Abbott}}]{Abbott2016}
\bibinfo{author}{\bibfnamefont{X.}~\bibnamefont{Wang}},
  \bibinfo{author}{\bibfnamefont{D.~S.} \bibnamefont{Miller}},
  \bibinfo{author}{\bibfnamefont{E.}~\bibnamefont{Bukusoglu}},
  \bibinfo{author}{\bibfnamefont{J.~J.} \bibnamefont{de~Pablo}},
  \bibnamefont{and} \bibinfo{author}{\bibfnamefont{N.~L.}
  \bibnamefont{Abbott}}, \bibinfo{journal}{Nature Mater.}
  \textbf{\bibinfo{volume}{15}}, \bibinfo{pages}{106} (\bibinfo{year}{2016}).

\bibitem[{\citenamefont{Kleman}(1983)}]{Kleman1983}
\bibinfo{author}{\bibfnamefont{M.}~\bibnamefont{Kleman}},
  \emph{\bibinfo{title}{Points, Lines and Walls}} (\bibinfo{publisher}{Wiley},
  \bibinfo{address}{New York}, \bibinfo{year}{1983}).

\bibitem[{\citenamefont{H\"anschke et~al.}(2017)\citenamefont{H\"anschke,
  Danilewsky, Helfen, Hamann, and Baumbach}}]{Hanschke2017}
\bibinfo{author}{\bibfnamefont{D.}~\bibnamefont{H\"anschke}},
  \bibinfo{author}{\bibfnamefont{A.}~\bibnamefont{Danilewsky}},
  \bibinfo{author}{\bibfnamefont{L.}~\bibnamefont{Helfen}},
  \bibinfo{author}{\bibfnamefont{E.}~\bibnamefont{Hamann}}, \bibnamefont{and}
  \bibinfo{author}{\bibfnamefont{T.}~\bibnamefont{Baumbach}},
  \bibinfo{journal}{Phys. Rev. Lett.} \textbf{\bibinfo{volume}{119}},
  \bibinfo{pages}{215504} (\bibinfo{year}{2017}).

\bibitem[{\citenamefont{Friedel}(1964)}]{Friedel1964}
\bibinfo{author}{\bibfnamefont{J.}~\bibnamefont{Friedel}},
  \emph{\bibinfo{title}{Dislocations}} (\bibinfo{publisher}{Pergamon Press},
  \bibinfo{address}{Oxford}, \bibinfo{year}{1964}).

\bibitem[{\citenamefont{Ho{\l}yst and Oswald}(1995)}]{Holyst1995}
\bibinfo{author}{\bibfnamefont{R.}~\bibnamefont{Ho{\l}yst}} \bibnamefont{and}
  \bibinfo{author}{\bibfnamefont{P.}~\bibnamefont{Oswald}},
  \bibinfo{journal}{Int. J. Mod. Phys. B} \textbf{\bibinfo{volume}{9}},
  \bibinfo{pages}{1515} (\bibinfo{year}{1995}).

\bibitem[{\citenamefont{Pieranski}(2016)}]{Pieranski2016}
\bibinfo{author}{\bibfnamefont{P.}~\bibnamefont{Pieranski}},
  \bibinfo{journal}{C. R. Physique} \textbf{\bibinfo{volume}{17}},
  \bibinfo{pages}{242} (\bibinfo{year}{2016}).

\bibitem[{\citenamefont{Meyer et~al.}(1978)\citenamefont{Meyer, Stebler, and
  Lagerwall}}]{Meyer1978}
\bibinfo{author}{\bibfnamefont{R.~B.} \bibnamefont{Meyer}},
  \bibinfo{author}{\bibfnamefont{B.}~\bibnamefont{Stebler}}, \bibnamefont{and}
  \bibinfo{author}{\bibfnamefont{S.~T.} \bibnamefont{Lagerwall}},
  \bibinfo{journal}{Phys. Rev. Lett.} \textbf{\bibinfo{volume}{41}},
  \bibinfo{pages}{1393} (\bibinfo{year}{1978}).

\bibitem[{\citenamefont{Kamien and Lubensky}(1997)}]{Kamien1997}
\bibinfo{author}{\bibfnamefont{R.~D.} \bibnamefont{Kamien}} \bibnamefont{and}
  \bibinfo{author}{\bibfnamefont{T.~C.} \bibnamefont{Lubensky}},
  \bibinfo{journal}{J. Phys II France} \textbf{\bibinfo{volume}{7}},
  \bibinfo{pages}{157} (\bibinfo{year}{1997}).

\bibitem[{\citenamefont{Trivedi et~al.}(2012)\citenamefont{Trivedi, Klevets,
  Senyuk, Lee, and Smalyukh}}]{Trivedi2012}
\bibinfo{author}{\bibfnamefont{R.~P.} \bibnamefont{Trivedi}},
  \bibinfo{author}{\bibfnamefont{I.~I.} \bibnamefont{Klevets}},
  \bibinfo{author}{\bibfnamefont{B.}~\bibnamefont{Senyuk}},
  \bibinfo{author}{\bibfnamefont{T.}~\bibnamefont{Lee}}, \bibnamefont{and}
  \bibinfo{author}{\bibfnamefont{I.~I.} \bibnamefont{Smalyukh}},
  \bibinfo{journal}{Proc. Natl. Acad. Sci. USA} \textbf{\bibinfo{volume}{109}},
  \bibinfo{pages}{4744} (\bibinfo{year}{2012}).

\bibitem[{\citenamefont{Aharoni et~al.}(2017)\citenamefont{Aharoni, Machon, and
  Kamien}}]{Kamien2017}
\bibinfo{author}{\bibfnamefont{H.}~\bibnamefont{Aharoni}},
  \bibinfo{author}{\bibfnamefont{T.}~\bibnamefont{Machon}}, \bibnamefont{and}
  \bibinfo{author}{\bibfnamefont{R.~D.} \bibnamefont{Kamien}},
  \bibinfo{journal}{Phys. Rev. Lett.} \textbf{\bibinfo{volume}{118}},
  \bibinfo{pages}{257801} (\bibinfo{year}{2017}).

\bibitem[{\citenamefont{Sung et~al.}(2018)\citenamefont{Sung, de~la Cotte, and
  Grelet}}]{NatCom}
\bibinfo{author}{\bibfnamefont{B.}~\bibnamefont{Sung}},
  \bibinfo{author}{\bibfnamefont{A.}~\bibnamefont{de~la Cotte}},
  \bibnamefont{and} \bibinfo{author}{\bibfnamefont{E.}~\bibnamefont{Grelet}},
  \bibinfo{journal}{Nat. Commun.} \textbf{\bibinfo{volume}{9}},
  \bibinfo{pages}{1405} (\bibinfo{year}{2018}).

\bibitem[{\citenamefont{Chan and Webb}(1981)}]{Chan1981}
\bibinfo{author}{\bibfnamefont{W.~K.} \bibnamefont{Chan}} \bibnamefont{and}
  \bibinfo{author}{\bibfnamefont{W.~W.} \bibnamefont{Webb}},
  \bibinfo{journal}{J. Phys. Lett. (Paris)} \textbf{\bibinfo{volume}{42}},
  \bibinfo{pages}{1007} (\bibinfo{year}{1981}).

\bibitem[{\citenamefont{Maaloum et~al.}(1992)\citenamefont{Maaloum,
  Ausserr{\'e}, Chatenay, Coulon, and Gallot}}]{Maloum1992}
\bibinfo{author}{\bibfnamefont{M.}~\bibnamefont{Maaloum}},
  \bibinfo{author}{\bibfnamefont{D.}~\bibnamefont{Ausserr{\'e}}},
  \bibinfo{author}{\bibfnamefont{D.}~\bibnamefont{Chatenay}},
  \bibinfo{author}{\bibfnamefont{G.}~\bibnamefont{Coulon}}, \bibnamefont{and}
  \bibinfo{author}{\bibfnamefont{Y.}~\bibnamefont{Gallot}},
  \bibinfo{journal}{Phys. Rev. Lett.} \textbf{\bibinfo{volume}{68}},
  \bibinfo{pages}{1575} (\bibinfo{year}{1992}).

\bibitem[{\citenamefont{Zasadzinski}(1990)}]{Zasadzinski1990}
\bibinfo{author}{\bibfnamefont{J.~A.~N.} \bibnamefont{Zasadzinski}},
  \bibinfo{journal}{J. Phys. France} \textbf{\bibinfo{volume}{51}},
  \bibinfo{pages}{747} (\bibinfo{year}{1990}).

\bibitem[{\citenamefont{Zhang et~al.}(2015)\citenamefont{Zhang, Grubb, Seed,
  Sampson, Jakli, and Lavrentovich}}]{Zhang2015}
\bibinfo{author}{\bibfnamefont{C.}~\bibnamefont{Zhang}},
  \bibinfo{author}{\bibfnamefont{A.~M.} \bibnamefont{Grubb}},
  \bibinfo{author}{\bibfnamefont{A.~J.} \bibnamefont{Seed}},
  \bibinfo{author}{\bibfnamefont{P.}~\bibnamefont{Sampson}},
  \bibinfo{author}{\bibfnamefont{A.}~\bibnamefont{Jakli}}, \bibnamefont{and}
  \bibinfo{author}{\bibfnamefont{O.~D.} \bibnamefont{Lavrentovich}},
  \bibinfo{journal}{Phys. Rev. Lett.} \textbf{\bibinfo{volume}{115}},
  \bibinfo{pages}{087801} (\bibinfo{year}{2015}).

\bibitem[{\citenamefont{Ishikawa and Lavrentovich}(1999)}]{Ishikawa}
\bibinfo{author}{\bibfnamefont{T.}~\bibnamefont{Ishikawa}} \bibnamefont{and}
  \bibinfo{author}{\bibfnamefont{O.~D.} \bibnamefont{Lavrentovich}},
  \bibinfo{journal}{Phys. Rev. E} \textbf{\bibinfo{volume}{60}},
  \bibinfo{pages}{R5037} (\bibinfo{year}{1999}).

\bibitem[{\citenamefont{Smalyukh and Lavrentovich}(2002)}]{Smalyukh2002}
\bibinfo{author}{\bibfnamefont{I.~I.} \bibnamefont{Smalyukh}} \bibnamefont{and}
  \bibinfo{author}{\bibfnamefont{O.~D.} \bibnamefont{Lavrentovich}},
  \bibinfo{journal}{Phys. Rev. E} \textbf{\bibinfo{volume}{66}},
  \bibinfo{pages}{051703} (\bibinfo{year}{2002}).

\bibitem[{\citenamefont{Engstr\"om et~al.}(2011)\citenamefont{Engstr\"om,
  Trivedi, Persson, Goks\"or, Bertness, and Smalyukh}}]{Engstrom2011}
\bibinfo{author}{\bibfnamefont{D.}~\bibnamefont{Engstr\"om}},
  \bibinfo{author}{\bibfnamefont{R.~P.} \bibnamefont{Trivedi}},
  \bibinfo{author}{\bibfnamefont{M.}~\bibnamefont{Persson}},
  \bibinfo{author}{\bibfnamefont{M.}~\bibnamefont{Goks\"or}},
  \bibinfo{author}{\bibfnamefont{K.~A.} \bibnamefont{Bertness}},
  \bibnamefont{and} \bibinfo{author}{\bibfnamefont{I.~I.}
  \bibnamefont{Smalyukh}}, \bibinfo{journal}{Soft Matter}
  \textbf{\bibinfo{volume}{7}}, \bibinfo{pages}{6304} (\bibinfo{year}{2011}).

\bibitem[{\citenamefont{Zhang et~al.}(2013)\citenamefont{Zhang, Sadashiva,
  Lavrentovich, and J\'akli}}]{Zhang2013}
\bibinfo{author}{\bibfnamefont{C.}~\bibnamefont{Zhang}},
  \bibinfo{author}{\bibfnamefont{B.~K.} \bibnamefont{Sadashiva}},
  \bibinfo{author}{\bibfnamefont{O.~D.} \bibnamefont{Lavrentovich}},
  \bibnamefont{and} \bibinfo{author}{\bibfnamefont{A.}~\bibnamefont{J\'akli}},
  \bibinfo{journal}{Liq. Cryst.} \textbf{\bibinfo{volume}{40}},
  \bibinfo{pages}{1636} (\bibinfo{year}{2013}).

\bibitem[{\citenamefont{de~la Cotte et~al.}(2017)\citenamefont{de~la Cotte, Wu,
  Trevisan, Repula, and Grelet}}]{ACSnano}
\bibinfo{author}{\bibfnamefont{A.}~\bibnamefont{de~la Cotte}},
  \bibinfo{author}{\bibfnamefont{C.}~\bibnamefont{Wu}},
  \bibinfo{author}{\bibfnamefont{M.}~\bibnamefont{Trevisan}},
  \bibinfo{author}{\bibfnamefont{A.}~\bibnamefont{Repula}}, \bibnamefont{and}
  \bibinfo{author}{\bibfnamefont{E.}~\bibnamefont{Grelet}},
  \bibinfo{journal}{ACS Nano} \textbf{\bibinfo{volume}{11}},
  \bibinfo{pages}{10616} (\bibinfo{year}{2017}).

\bibitem[{\citenamefont{Dogic and Fraden}(2006)}]{Dogic2006}
\bibinfo{author}{\bibfnamefont{Z.}~\bibnamefont{Dogic}} \bibnamefont{and}
  \bibinfo{author}{\bibfnamefont{S.}~\bibnamefont{Fraden}},
  \bibinfo{journal}{Curr. Opin. Colloid Interface Sci.}
  \textbf{\bibinfo{volume}{11}}, \bibinfo{pages}{47} (\bibinfo{year}{2006}).

\bibitem[{\citenamefont{Grelet}(2014)}]{Grelet2014}
\bibinfo{author}{\bibfnamefont{E.}~\bibnamefont{Grelet}},
  \bibinfo{journal}{Phys. Rev. X} \textbf{\bibinfo{volume}{4}},
  \bibinfo{pages}{021053} (\bibinfo{year}{2014}).

\bibitem[{\citenamefont{Brener and Marchenko}(1999)}]{Brener99}
\bibinfo{author}{\bibfnamefont{E.~A.} \bibnamefont{Brener}} \bibnamefont{and}
  \bibinfo{author}{\bibfnamefont{V.~I.} \bibnamefont{Marchenko}},
  \bibinfo{journal}{Phys. Rev. E} \textbf{\bibinfo{volume}{59}},
  \bibinfo{pages}{R4752} (\bibinfo{year}{1999}).

\bibitem[{\citenamefont{de~Gennes}(1972)}]{DeGennes1972}
\bibinfo{author}{\bibfnamefont{P.~G.} \bibnamefont{de~Gennes}},
  \bibinfo{journal}{C. R. Hebd. S\'{e}ances Acad. Sci.}
  \textbf{\bibinfo{volume}{275B}}, \bibinfo{pages}{939} (\bibinfo{year}{1972}).

\bibitem[{\citenamefont{Barry and Dogic}(2010)}]{Barry2010}
\bibinfo{author}{\bibfnamefont{E.}~\bibnamefont{Barry}} \bibnamefont{and}
  \bibinfo{author}{\bibfnamefont{Z.}~\bibnamefont{Dogic}},
  \bibinfo{journal}{Proc. Natl. Acad. Sci. USA} \textbf{\bibinfo{volume}{107}},
  \bibinfo{pages}{10348} (\bibinfo{year}{2010}).

\bibitem[{\citenamefont{Dogic and Fraden}(2000)}]{Dogic2000}
\bibinfo{author}{\bibfnamefont{Z.}~\bibnamefont{Dogic}} \bibnamefont{and}
  \bibinfo{author}{\bibfnamefont{S.}~\bibnamefont{Fraden}},
  \bibinfo{journal}{Langmuir} \textbf{\bibinfo{volume}{16}},
  \bibinfo{pages}{7820} (\bibinfo{year}{2000}).

\bibitem[{\citenamefont{Shibahara et~al.}(2000)\citenamefont{Shibahara,
  Yamamoto, Takanishi, Ishikawa, Takezoe, and Tanaka}}]{Shibahara2000}
\bibinfo{author}{\bibfnamefont{S.}~\bibnamefont{Shibahara}},
  \bibinfo{author}{\bibfnamefont{J.}~\bibnamefont{Yamamoto}},
  \bibinfo{author}{\bibfnamefont{Y.}~\bibnamefont{Takanishi}},
  \bibinfo{author}{\bibfnamefont{K.}~\bibnamefont{Ishikawa}},
  \bibinfo{author}{\bibfnamefont{H.}~\bibnamefont{Takezoe}}, \bibnamefont{and}
  \bibinfo{author}{\bibfnamefont{H.}~\bibnamefont{Tanaka}},
  \bibinfo{journal}{Phys. Rev. Lett.} \textbf{\bibinfo{volume}{85}},
  \bibinfo{pages}{1670} (\bibinfo{year}{2000}).

\bibitem[{SM()}]{SM}
\bibinfo{note}{See Supplemental Material at http://link.aps.org/ for the width
  distribution of screw dislocation cores, a movie displaying the
  self-diffusion of tip-labeled rod-like particles within the defect core and
  the smectic bulk, with an example of rare hopping-type events.}

\bibitem[{\citenamefont{Pleiner}(1986)}]{Pleiner1986}
\bibinfo{author}{\bibfnamefont{H.}~\bibnamefont{Pleiner}},
  \bibinfo{journal}{Liq. Cryst.} \textbf{\bibinfo{volume}{1}},
  \bibinfo{pages}{197} (\bibinfo{year}{1986}).

\bibitem[{\citenamefont{Pleiner}(1988)}]{Pleiner1988}
\bibinfo{author}{\bibfnamefont{H.}~\bibnamefont{Pleiner}},
  \bibinfo{journal}{Liq. Cryst.} \textbf{\bibinfo{volume}{3}},
  \bibinfo{pages}{249} (\bibinfo{year}{1988}).

\bibitem[{\citenamefont{Matsumoto et~al.}(2012)\citenamefont{Matsumoto, Kamien,
  and Santangelo}}]{Matsumoto12}
\bibinfo{author}{\bibfnamefont{E.~A.} \bibnamefont{Matsumoto}},
  \bibinfo{author}{\bibfnamefont{R.~D.} \bibnamefont{Kamien}},
  \bibnamefont{and} \bibinfo{author}{\bibfnamefont{C.~D.}
  \bibnamefont{Santangelo}}, \bibinfo{journal}{Interface Focus}
  \textbf{\bibinfo{volume}{2}}, \bibinfo{pages}{617} (\bibinfo{year}{2012}).

\bibitem[{\citenamefont{Alvarez et~al.}(2017)\citenamefont{Alvarez, Lettinga,
  and Grelet}}]{Alvarez2017}
\bibinfo{author}{\bibfnamefont{L.}~\bibnamefont{Alvarez}},
  \bibinfo{author}{\bibfnamefont{M.~P.} \bibnamefont{Lettinga}},
  \bibnamefont{and} \bibinfo{author}{\bibfnamefont{E.}~\bibnamefont{Grelet}},
  \bibinfo{journal}{Phys. Rev. Lett.} \textbf{\bibinfo{volume}{118}},
  \bibinfo{pages}{178002} (\bibinfo{year}{2017}).

\bibitem[{\citenamefont{Kralj and Sluckin}(1993)}]{Kralj1993}
\bibinfo{author}{\bibfnamefont{S.}~\bibnamefont{Kralj}} \bibnamefont{and}
  \bibinfo{author}{\bibfnamefont{T.~J.} \bibnamefont{Sluckin}},
  \bibinfo{journal}{Phys. Rev. E} \textbf{\bibinfo{volume}{48}},
  \bibinfo{pages}{R3244} (\bibinfo{year}{1993}).

\bibitem[{\citenamefont{Kralj and Sluckin}(1995)}]{Kralj1995}
\bibinfo{author}{\bibfnamefont{S.}~\bibnamefont{Kralj}} \bibnamefont{and}
  \bibinfo{author}{\bibfnamefont{T.~J.} \bibnamefont{Sluckin}},
  \bibinfo{journal}{Liq. Cryst.} \textbf{\bibinfo{volume}{18}},
  \bibinfo{pages}{887} (\bibinfo{year}{1995}).

\end{thebibliography}

\clearpage
	
\end{document}